\documentclass[twocolumn]{aastex63}
\usepackage{graphicx}
\usepackage{ulem}
\usepackage{cellspace}
\usepackage{amssymb}
\usepackage{amsmath}
\usepackage{longtable}
\usepackage{tabu}
\usepackage{color}
\usepackage{etoolbox}
\usepackage{supertabular}
\usepackage{savesym}
\savesymbol{tablenum}
\usepackage{siunitx}
\restoresymbol{SIX}{tablenum}
\usepackage{dcolumn}
\usepackage{ wasysym }

\newcommand\Tstrut{\rule{0pt}{2.9ex}}       
\newcommand\Bstrut{\rule[-1.3ex]{0pt}{0pt}} 
\newcommand\TBstrut{\Tstrut\Bstrut}

\newcommand{\paramtotaltemplates}{\num{7.9e+17}}
\newcommand{\paramtotalWUsmillions}{\num{8}}
\newcommand{\paramNCandTotal}{\num{6.0e+10}}
\newcommand{\paramWUtotaltemplates}{\num{98277129500}}

\newcommand{\paramWUcputimeHours}{8}
\newcommand{\paramNCand}{\num{7500}}
\newcommand{\paramNfinWU}{\num{14970}}
\newcommand{\paramNfdotinWU}{\num{8855}}
\newcommand{\paramfmax}{585.15}
\newcommand{\paramfmin}{20.0}
\newcommand{\paramfdotmin}{\num{-2.6e-9}}
\newcommand{\paramfdotmax}{\num{2.6e-10}}
\newcommand{\paramTref}{1177858472.0}
\newcommand{\NDisturbedBands}{\num{     273}}

\newcommand{\GPSstart}{1167983370}
\newcommand{\GPStend}{1187731774}
\newcommand{\GPSstartDate}{Jan 09 2017}
\newcommand{\GPStendDate}{Aug 25 2017}

\newcommand{\meanMismatchFUZero}{\num{0.5}}
\newcommand{\meanMismatchFUZeroR}{\num{0.3}}
\newcommand{\meanMismatchFUOne}{\num{0.09}}
\newcommand{\meanMismatchFUOneR}{\num{0.002}}
\newcommand{\meanMismatchFUTwo}{\num{0.001}}
\newcommand{\meanMismatchFUThree}{\num{0.001}}
\newcommand{\meanMismatchFUFour}{\num{0.0002}}
\newcommand{\meanMismatchFUFive}{\num{0.0001}}
\newcommand{\meanMismatchFUSix}{\num{0.0007}}

\newcommand{\TcohFUZero}{\num{60}}
\newcommand{\TcohFUZeroR}{\num{60}}
\newcommand{\TcohFUOne}{\num{126}}
\newcommand{\TcohFUOneR}{\num{126}}
\newcommand{\TcohFUTwo}{\num{250}}
\newcommand{\TcohFUThree}{\num{500}}
\newcommand{\TcohFUFour}{\num{1000}}
\newcommand{\TcohFUFive}{\num{1563}}
\newcommand{\TcohFUSix}{\approx\num{5486}}

\newcommand{\NSegFUZero}{64}
\newcommand{\NSegFUZeroR}{64}
\newcommand{\NSegFUOne}{29}
\newcommand{\NSegFUOneR}{29}
\newcommand{\NSegFUTwo}{14}
\newcommand{\NSegFUThree}{7}
\newcommand{\NSegFUFour}{2}
\newcommand{\NSegFUFive}{2}
\newcommand{\NSegFUSix}{1}

\newcommand{\dfreqmuHzFUZero}{\num{3.34}}

\newcommand{\dfdotfgTenfHzFUZero}{\num{32.7479}}

\newcommand{\mskyFUZero}{\num{8.0e-3}}

\newcommand{\dfreqmuHzFUZeroR}{\num{3.34}}

\newcommand{\dfdotfgTenfHzFUZeroR}{\num{20}}
\newcommand{\mskyFUZeroR}{\num{5.0e-04}}

\newcommand{\dfreqmuHzFUOne}{\num{1}}

\newcommand{\dfdotfgTenfHzFUOne}{\num{2}}
\newcommand{\mskyFUOne}{\num{1.0e-05}}

\newcommand{\dfreqmuHzFUOneR}{\num{0.19}}

\newcommand{\dfdotfgTenfHzFUOneR}{\num{2}}
\newcommand{\mskyFUOneR}{\num{1.0e-07}}

\newcommand{\dfreqmuHzFUTwo}{\num{0.025}}

\newcommand{\dfdotfgTenfHzFUTwo}{\num{2}}
\newcommand{\mskyFUTwo}{\num{2.5e-08}}

\newcommand{\dfreqmuHzFUThree}{\num{0.01}}

\newcommand{\dfdotfgTenfHzFUThree}{\num{1}}
\newcommand{\mskyFUThree}{\num{1.0e-08}}

\newcommand{\dfreqmuHzFUFour}{\num{0.001}}

\newcommand{\dfdotfgTenfHzFUFour}{\num{0.1}}
\newcommand{\mskyFUFour}{\num{1.0e-09}}

\newcommand{\dfreqmuHzFUFive}{\num{0.001}}

\newcommand{\dfdotfgTenfHzFUFive}{\num{0.1}}
\newcommand{\mskyFUFive}{\num{5.0e-10}}

\newcommand{\dfreqmuHzFUSix}{\num{0.001}}

\newcommand{\dfdotfgTenfHzFUSix}{\num{0.1}}
\newcommand{\mskyFUSix}{\num{1.0e-10}}

\newcommand{\CRfdotFUZero}{\num{1.2e-10 }}
\newcommand{\CRfreqmuHzFUZero}{\num{850.}}

\newcommand{\CRskyFUZero}{\num{5.0}}

\newcommand{\CRfdotFUZeroR}{\num{2.0e-11 }}
\newcommand{\CRfreqmuHzFUZeroR}{\num{130.}}

\newcommand{\CRskyFUZeroR}{\num{0.75}}

\newcommand{\CRfdotFUOne}{\num{2.0e-12 }}
\newcommand{\CRfreqmuHzFUOne}{\num{10.}}

\newcommand{\CRskyFUOne}{\num{0.1}}

\newcommand{\CRfdotFUOneR}{\num{3.2e-13 }}
\newcommand{\CRfreqmuHzFUOneR}{\num{0.4}}

\newcommand{\CRskyFUOneR}{\num{0.02}}

\newcommand{\CRfdotFUTwo}{\num{1.45e-13 }}
\newcommand{\CRfreqmuHzFUTwo}{\num{0.17}}

\newcommand{\CRskyFUTwo}{\num{0.008}}

\newcommand{\CRfdotFUThree}{\num{6.4e-14 }}
\newcommand{\CRfreqmuHzFUThree}{\num{0.067}}

\newcommand{\CRskyFUThree}{\num{0.0037}}

\newcommand{\CRfdotFUFour}{\num{8.0e-14 }}
\newcommand{\CRfreqmuHzFUFour}{\num{0.05}}

\newcommand{\CRskyFUFour}{\num{0.005}}

\newcommand{\CRfdotFUFive}{\num{4.25e-14 }}
\newcommand{\CRfreqmuHzFUFive}{\num{0.0325}}

\newcommand{\CRskyFUFive}{\num{0.0025}}

\newcommand{\rVetoFUZero}{-}

\newcommand{\rVetoFUZeroR}{0.75}

\newcommand{\rVetoFUOne}{1.99}

\newcommand{\rVetoFUOneR}{2.2}

\newcommand{\rVetoFUTwo}{4.3}

\newcommand{\rVetoFUThree}{6.0}

\newcommand{\rVetoFUFour}{10.0}

\newcommand{\rVetoFUFive}{15.0}

\newcommand{\rVetoFUSix}{50.0}

\newcommand{\NCandFUZero}{\num{350145}}
\newcommand{\NCandFUZeroR}{\num{101001}}
\newcommand{\NCandFUOne}{\num{11915}}
\newcommand{\NCandFUOneR}{\num{6128}}
\newcommand{\NCandFUTwo}{\num{33}}
\newcommand{\NCandFUThree}{\num{21}}
\newcommand{\NCandFUFour}{\num{18}}
\newcommand{\NCandFUFive}{\num{8}}
\newcommand{\NCandDisturbedFUZero}{\num{    1352}}

\newcommand{\mostStringentUL}{\num{1.3e-25}}
\newcommand{\mostStringentULFreq}{\num{163.0}}
\newcommand{\minSensitivityDepth}{\num{49}}
\newcommand{\maxSensitivityDepth}{\num{56}}

\def\Tcoh{T_{\textrm{\mbox{\tiny{coh}}}}}

\def\min{\textrm{\mbox{\tiny{min}}}}
\def\max{\textrm{\mbox{\tiny{max}}}}

\def\EaH{Einstein@Home}

\renewcommand{\sec}{\mathrm{s}}

\newcommand{\avgSeg}[1]{\overline{#1}}			

\newcommand{\Freq}{f}
\newcommand{\fdot}{{\dot{\Freq}}}

\newcommand{\M}{\mathcal{M}}

\newcommand{\Gauss}{\mathrm{\MakeUppercase{G}}}
\newcommand{\Signal}{{\mathrm{\MakeUppercase{S}}}}
\newcommand{\Line}{{\mathrm{\MakeUppercase{L}}}}
\newcommand{\Transient}{{\mathrm{t\MakeUppercase{L}}}}

\newcommand{\NoisetL}{{\Gauss\Line\Transient}}

\providecommand{\sc}[1]{\widehat{#1}}
\renewcommand{\sc}[1]{\widehat{#1}}

\newcommand{\BSNtsc}{{\hat\beta}_{{\Signal/\NoisetL}}}	

\newcommand{\F}{\mathcal{F}}		

\newcommand{\avF}{\avgSeg{\F}}

\newcommand{\Nseg}{{N_{\mathrm{seg}}}}

\renewcommand{\csc}{$\sc{\textrm{c}}$}

\newcommand{\D}{{\mathcal{D}}}

\shorttitle{O2 Einstein@Home all-sky search for continuous gravitational waves}
\shortauthors{Steltner, Papa et al.}

\begin{document}

\title{Einstein@Home all-sky search for continuous gravitational waves in LIGO O2 public data}

\correspondingauthor{B. Steltner}
\email{benjamin.steltner@aei.mpg.de}

\correspondingauthor{M.A. Papa}
\email{maria.alessandra.papa@aei.mpg.de}

\author[0000-0003-1833-5493]{B. Steltner}
\affiliation{Max Planck Institute for Gravitational Physics (Albert Einstein Institute), Callinstrasse 38, 30167 Hannover, Germany}
\affiliation{Leibniz Universit\"at Hannover, D-30167 Hannover, Germany}

\author[0000-0002-1007-5298]{M. A. Papa}
\affiliation{Max Planck Institute for Gravitational Physics (Albert Einstein Institute), Callinstrasse 38, 30167 Hannover, Germany}
\affiliation{University of Wisconsin Milwaukee, 3135 N Maryland Ave, Milwaukee, WI 53211, USA}
\affiliation{Leibniz Universit\"at Hannover, D-30167 Hannover, Germany}

\author{H.-B. Eggenstein}
\affiliation{Max Planck Institute for Gravitational Physics (Albert Einstein Institute), Callinstrasse 38, 30167 Hannover, Germany}
\affiliation{Leibniz Universit\"at Hannover, D-30167 Hannover, Germany}

\author{B. Allen}
\affiliation{Max Planck Institute for Gravitational Physics (Albert Einstein Institute), Callinstrasse 38, 30167 Hannover, Germany}
\affiliation{University of Wisconsin Milwaukee, 3135 N Maryland Ave, Milwaukee, WI 53211, USA}
\affiliation{Leibniz Universit\"at Hannover, D-30167 Hannover, Germany}

\author{V. Dergachev}
\affiliation{Max Planck Institute for Gravitational Physics (Albert Einstein Institute), Callinstrasse 38, 30167 Hannover, Germany}
\affiliation{Leibniz Universit\"at Hannover, D-30167 Hannover, Germany}

\author{R. Prix}
\affiliation{Max Planck Institute for Gravitational Physics (Albert Einstein Institute), Callinstrasse 38, 30167 Hannover, Germany}
\affiliation{Leibniz Universit\"at Hannover, D-30167 Hannover, Germany}

\author{B. Machenschalk}
\affiliation{Max Planck Institute for Gravitational Physics (Albert Einstein Institute), Callinstrasse 38, 30167 Hannover, Germany}
\affiliation{Leibniz Universit\"at Hannover, D-30167 Hannover, Germany}

\author{S. Walsh}
\affiliation{Max Planck Institute for Gravitational Physics (Albert Einstein Institute), Callinstrasse 38, 30167 Hannover, Germany}
\affiliation{University of Wisconsin Milwaukee, 3135 N Maryland Ave, Milwaukee, WI 53211, USA}
\affiliation{Leibniz Universit\"at Hannover, D-30167 Hannover, Germany}

\author{S. J. Zhu}
\affiliation{Max Planck Institute for Gravitational Physics (Albert Einstein Institute), Callinstrasse 38, 30167 Hannover, Germany}
\affiliation{Leibniz Universit\"at Hannover, D-30167 Hannover, Germany}
\affiliation{DESY, D-15738 Zeuthen, Germany}

\author{S. Kwang}
\affiliation{University of Wisconsin Milwaukee, 3135 N Maryland Ave, Milwaukee, WI 53211, USA}

\begin{abstract}
We conduct an all-sky search for continuous gravitational waves in the LIGO O2 data from the Hanford and Livingston detectors. We search for nearly-monochromatic signals with frequency $\paramfmin\ \textrm{Hz} \leq f \leq \paramfmax $ Hz 
and spin-down $\paramfdotmin\ \textrm{Hz/s} \leq \fdot \leq \paramfdotmax $ Hz/s. We deploy the search on the \EaH\ volunteer-computing project and follow-up the waveforms associated with the most significant results with eight further search-stages, reaching the best sensitivity ever achieved by an all-sky survey up to 500 Hz. Six of the inspected waveforms pass all the stages but they are all associated with hardware-injections, which are fake signals simulated at the LIGO detector for validation purposes. We recover all these fake signals with consistent parameters. No other waveform survives, so we find no evidence of a continuous gravitational wave signal at the detectability level of our search. We constrain the $h_0$ amplitude of continuous gravitational waves at the detector as a function of the signal frequency, in half-Hz bins. The most constraining upper limit at \mostStringentULFreq\ Hz is $h_0=\mostStringentUL$, at the 90\% confidence level.
Our results exclude neutron stars rotating faster than 5 ms 
with equatorial ellipticities larger than $10^{-7}$ closer than 100 pc.
These are deformations that neutron star crusts could easily support, according to some models.

\end{abstract}

\keywords{continuous gravitational waves, neutron stars}

\section{Introduction}

Continuous gravitational waves are expected in a variety of astrophysical scenarios: from rotating neutrons stars if they present some sort of asymmetry with respect to their rotation axis or through the excitation of unstable r-modes \citep{Lasky:2015uia,Owen:1998xg}; from the fast inspiral of dark-matter objects \citep{Horowitz:2019aim,Horowitz:2019pru}; through super-radiant emission of axion-like particles around black holes \citep{Arvanitaki:2014wva,Zhu:2020tht}.

The expected gravitational wave amplitude at the Earth is several orders of magnitude smaller than that of signals from compact binary inspirals, but because the signal is long-lasting one can integrate it over many months and increase the signal-to-noise ratio (SNR) very significantly. 

The most challenging searches for this type of signal are the all-sky surveys, where one looks for a signal from a source that is not known. 
The main challenge of these searches is that the number of waveforms that can be resolved over months of observation is very large, and so the sensitivity of the search is limited by its computational cost. 

In this paper we present the results from an all-sky search for continuous gravitational wave signals
with frequency $f$ between $\paramfmin$ Hz and  $\paramfmax$ Hz and spin-down $\paramfdotmin\ \textrm{Hz/s} \leq \fdot \leq \paramfdotmax $ Hz/s, carried out thanks to the computing power donated by the volunteers of the \EaH\ project. 

The results from the \EaH\ search are further processed using a hierarchy of eight follow-up searches, similarly to what previously done for recent \EaH\ searches \citep{Abbott:2017pqa, Ming:2019xse, Papa:2020vfz}. 

We use LIGO O2 public data \citep{Abbott:2019ebz,Vallisneri:2014vxa, o2_data} and, thanks to a much longer coherent-search baseline,  achieve a significantly higher sensitivity than the LIGO Collaboration O2 results in the same frequency range \citep{Pisarski:2019vxw, Palomba:2019vxe}. Our results complement those of the high-frequency Falcon search \citep{Dergachev:2020fli}, which cover the range from 500 to 1700 Hz. 

The plan of the paper is the following: we introduce the signal model and generalities about the search in Sections \ref{sec:signal} and \ref{sec:SearchGeneral}, respectively. In Sections \ref{sec:stage0} and 
\ref{sec:followups} we detail the \EaH\ search and the follow-up searches. Constraints on the gravitational wave amplitude and on the ellipticity of neutron stars are obtained in Section \ref{sec:ULs}, and conclusions are drawn in Section \ref{sec:conclusions}.

\section{The signal}
\label{sec:signal}

The search described in this paper targets nearly monochromatic gravitational wave signals of the form described for example in Section II of  \cite{Jaranowski:1998qm}.  At the output of a gravitational wave detector the signal has the form
\begin{equation}
h(t)=F_+ (\alpha,\delta,\psi ;t) h_+ (t) + F_\times (\alpha,\delta,\psi; t) h_\times(t).
\label{eq:signal}
\end{equation}
$F_+ (\alpha,\delta,\psi;t)$ and
$F_\times(\alpha,\delta,\psi;t)$ are the detector beam-pattern functions for the ``+" and ``$\times$" polarizations, $(\alpha,\delta)$ the
right-ascension and declination of the source, $\psi$ the polarization angle
and $t$ the time at the detector. The waveforms $h_+ (t)$ and $h_\times (t)$ take the form
\begin{eqnarray}
h_+ (t)  =  A_+ \cos \Phi(t) \nonumber \\
h_\times (t)  =  A_\times \sin \Phi(t),
\label{eq:monochromatic}
\end{eqnarray}
with the ``+" and ``$\times$" amplitudes 
\begin{eqnarray}
A_+  & = & {1\over 2} h_0 (1+\cos^2\iota) \nonumber \\
A_\times & = &  h_0  \cos\iota. 
\label{eq:amplitudes}
\end{eqnarray}
The angle between the total angular momentum of the star and the line of sight is $0\leq \iota \leq \pi$ and 
$h_0\geq 0$ is the intrinsic gravitational wave amplitude. 
$\Phi(t)$ of Eq.~\ref{eq:monochromatic} is the phase of the gravitational wave signal at time
$t$. If $\tau_{\mathrm{SSB}}$ is the arrival time of the wave with phase $\Phi(t)$ at the solar system barycenter, then $\Phi(t)=\Phi(\tau_{\mathrm{SSB}}(t))$. The gravitational wave phase as function of $\tau_{\mathrm{SSB}}$ is assumed to be 
\begin{multline}
\label{eq:phiSSB}
\Phi(\tau_{\mathrm{SSB}}) = \Phi_0 + 2\pi [ f(\tau_{\mathrm{SSB}}-{\tau_0}_{\mathrm{SSB}})  +
\\ {1\over 2} \dot{f} (\tau_{\mathrm{SSB}}-{\tau_0}_{\mathrm{SSB}})^2 ].
\end{multline}
We take ${\tau_0}_{\mathrm{SSB}}=$ \paramTref\ (TDB in GPS seconds) as a reference time.

\section{Generalities of the searches}
\label{sec:SearchGeneral}

\subsection{The data}
\label{subsec:data}

We use LIGO O2 public data from the Hanford (LHO) and the Livingston (LLO) detectors between GPS time \GPSstart\ (\GPSstartDate) and \GPStend\ (\GPStendDate). This data has been treated to remove spurious noise due to the LIGO laser beam jitter, calibration lines and the mains power lines \citep{Davis:2018yrz}. 

We additionally remove very loud short-duration glitches \citep{steltner_gating} and substitute Gaussian noise at frequency bins affected by line contamination \citep{Covas:2018oik}. This is a procedure common to all \EaH\ searches and it prevents spectral contamination from spreading to many nearby signal frequencies. The list of cleaned frequency bins can be found in \citet[and Suppl. Mat.]{O2AS-AEI}.

\begin{figure}[h!tbp]
   \includegraphics[width=\columnwidth]{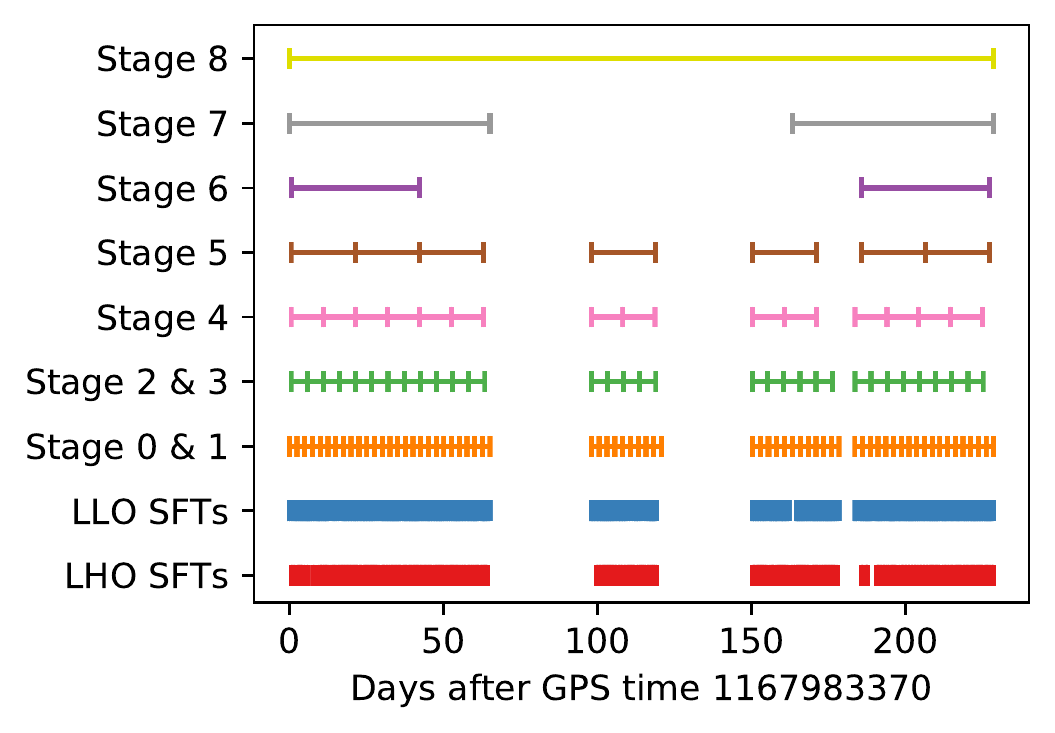}
\caption{Segmentation of the data used for the \EaH\ search and the follow-up stages. The lower two bars show the input SFTs. The first gap in the data -- starting at $\lessapprox$ 70 days -- is due to spectral contamination in LHO, based on which we decided to exclude this period from the analysis. The second large gap -- starting at $\approx$ 120 days -- is due to an interruption of the science run for detector commissioning.} 
\label{fig:NsegSegmentation}
\end{figure}

As is customary, the input to our searches is in the form of Short time-baseline (30 minutes) Fourier Transforms (SFTs). These are grouped in segments of variable duration, that correspond to the coherent time baselines of the various searches, as shown in Figure \ref{fig:NsegSegmentation}. 

\subsection{The detection statistics}
\label{subsec:search}

For each search we partition the data in $\Nseg$ segments, with each segment spanning a duration $\Tcoh$. The data of both detectors from each segment $i$ are  combined coherently to construct a matched-filter detection statistic, the $\F$-statistic \citep{Cutler:2005hc}. The $\F$-statistic depends on the template waveform that is being tested for consistency with the data. The $\F$-statistic at a given template point is the log-likelihood ratio of the data containing Gaussian noise plus a signal with the shape given by the template, to the data being purely Gaussian noise. 

The $\F$-statistic values are summed, one per segment ($\F_i$), and, after dividing by $\Nseg$, this yields our core detection statistic \citep{Pletsch:2009uu,Pletsch:2010xb}: 
\begin{equation}
\label{eq:avF}
\avF:={1\over\Nseg} \sum_{i=1}^{\Nseg} \F_i.
\end{equation}
The $\avF$ is the average of $\F$ over segments, in general computed at different templates for every segment. The resulting $\avF$ is an approximation to the detection statistic at some template ``in between'' the ones used to compute the single-segment $\F_i$. In fact these ``in-between" templates constitute a finer grid based on which the summations of Eq.~\ref{eq:avF} are performed. 

The most significant \EaH\ results are saved in the top-list that the volunteer computer (host) returns to the \EaH\ server. For these results the host also re-computes $\avF$ at the exact fine-grid template point. We indicate the re-computed statistic with a subscript ``$r$", as for example in $\avF_r$. 

In Gaussian noise $\Nseg\times 2\avF$ follows a chi-squared distribution with $4\Nseg$ degrees of freedom, $\chi^2_{4\Nseg}(\rho^2)$. The non-centrality parameter $\rho^2\propto h_0^2 T_{\rm{data}}/S_h$, where $T_{\rm{data}}$ is the duration of time for which data is available and $S_h$ is the strain power spectral density of the noise. The expected SNR squared is equal to $\rho^2$ \citep{Jaranowski:1998qm}. For simplicity, in the rest of the paper, when we refer to the SNR we mean SNR squared.

If the noise contains some coherent instrumental or environmental signal, it is very likely that for some of the templates the distribution of $\avF$ will have a non-zero non-centrality parameter, even though there is no astrophysical signal. The reason is that in this case the data looks more like a noise+signal than like pure Gaussian noise.

It is possible to identify a non-astrophysical signal if it presents features that distinguish it from the astrophysical signals that the search is targeting, for example if it is present only in one of the two detectors, or if it is present only for part of the observation time. In the past we have used these signatures to construct ad-hoc vetoes, such as the $\F$-stat consistency veto \citep{Aasi:2012fw} and the permanence veto \citep{Behnke:2014tma,Aasi:2013jya}.  These vetoes are still widely used although with different names: the ``single interferometer veto" in \citet{Sun:2019mqb,Jones:2020htx} and the ``persistency veto" of \citet{Pisarski:2019vxw,Astone:2014esa}.

We incorporated the ideas of the  $\F$-stat consistency veto and of the permanence veto in the design of a new detection statistic, $\BSNtsc$. The new detection statistic is an odds ratio that tests the signal hypothesis against a noise model, which in addition to Gaussian noise also includes single-detector continuous or transient spectral lines \citep{Keitel:2013wga,Keitel:2015ova}. The subscript ``L'' in $\BSNtsc$ stands for line, ``G" for Gaussian and ``tL" for transient-line. We use this detection statistic to rank the \EaH\ results. In this way we limit the number of results that make it in the top-list but that would later be  discarded by the vetoes. This frees up space on the top list for other, more interesting, results.

\subsection{The search grids}
\label{subsec:grids}
For a rotating isolated neutron star, the template waveform is defined by the signal frequency, the spin-down and the source sky-position. The range searched in each of these variables is gridded in such a way that the the fractional loss in SNR, or mismatch, due to a signal falling in-between grid-points is on average \meanMismatchFUZero. 

\begin{figure}[h!tbp]
    \includegraphics[width=\columnwidth]{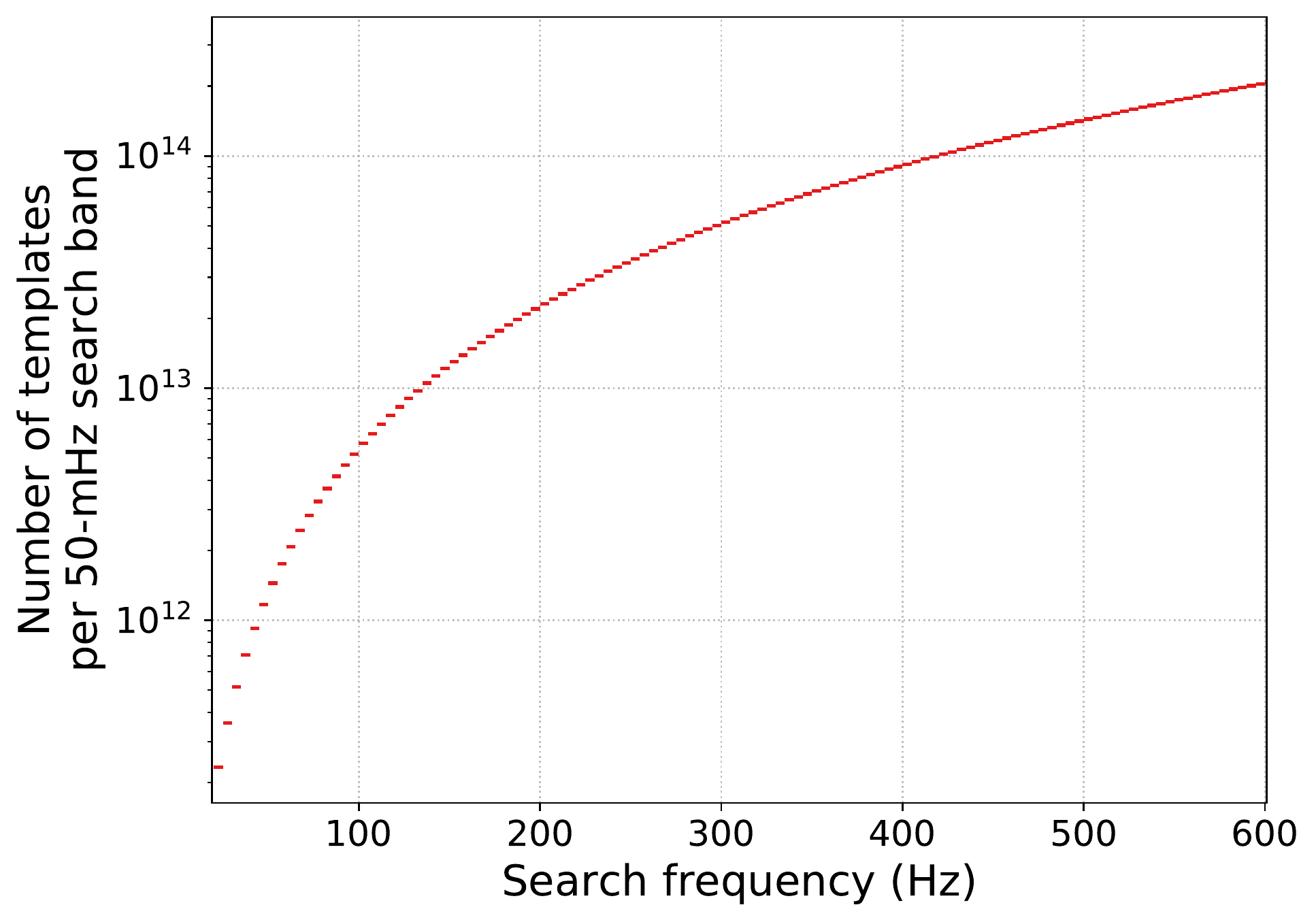}
\caption{Number of searched templates per 50-mHz band as a function of frequency. The sky resolution increases with frequency causing the increase in the number of templates. The number of templates in frequency and spindown is 
$\paramNfinWU$ and $\paramNfdotinWU$ respectively. 
}  
\label{fig:NumberOfTemplatesIn50mHz}
\end{figure}

The grids in frequency and spin-down are each described by a single parameter, the grid spacing, which is constant over the search range. The sky grid is approximately uniform on the celestial sphere orthogonally projected on the ecliptic plane. The tiling is an hexagonal covering of the unit circle with hexagon edge length $d$:
\begin{equation}
d(m_{\text{sky}})={1\over f}
{
{\sqrt{ m_{\text{sky}} } }
\over {\pi \tau_{E}}
} ,
\label{eq:skyGridSpacing}
\end{equation}
with $\tau_{E}\simeq0.021$ s being half of the light travel-time across the Earth and $m_{\text{sky}}$ a constant which controls the resolution of the sky grid. The sky-grids are constant over 5\,Hz bands and the spacings are the ones associated through Eq.~\ref{eq:skyGridSpacing} to the highest frequency in each 5\,Hz. The resulting number of templates used to search 50-mHz bands as a function of frequency is shown in Fig. \ref{fig:NumberOfTemplatesIn50mHz}. The grid spacings and $m_{\text{sky}}$ are given in Table \ref{tab:FUtable}. 

\subsection{The Monte Carlos and the assumed signal population}
\label{sub:MCs}

The loss in signal-to-noise ratio $\mu(\vec\lambda_0)$ due to the parameters $\vec\lambda_0$ of a signal not perfectly matching the parameters $\vec\lambda_0 \pm \Delta\vec\lambda$ of the template can be described by a quadratic form, as long as the signal and the template parameters are fairly close, i.e. as long as $\Delta\vec\lambda$ is small:
\begin{equation}
\mu(\vec\lambda_0)=g_{ij}(\vec\lambda_0) \Delta\lambda^i  \Delta\lambda^j.
\label{eq:metric}
\end{equation}
The metric $g_{ij}$ for the search at hand can be estimated, at least numerically. 

Setting up a search is a matter of deciding what loss in SNR one is willing to accept (fixing the mismatch), picking a tiling method and setting up a grid accordingly. Once the search grid is established, one can determine the computational cost of the search. If that is found to be too high, one must decide whether to reduce the $T_{coh}$ or to increase the mismatch, and repeat the procedure. Ultimately the best operating point is a compromise between computational cost and sensitivity. 

It turns out that for the first stages of all-sky surveys with $T_{coh}$ of at least several hours, the optimal grids are typically ones with spacings $\gg$ the ones at which the metric approximation of Eq.~\ref{eq:metric} holds. In particular we find that the metric mismatch overestimates the actual mismatch. This is good because it means that in order to achieve a certain maximum mismatch level, we need fewer templates than what the metric predicts. On the other hand it means that we cannot predict the mismatch analytically using Eq.~\ref{eq:metric}. Instead we must resort to simulating signals, searching for them with a given grid and measuring the loss in SNR with respect to a perfectly matched template. And we have to do this many times to probe different signals ($\vec\lambda_0$ values) and different random offsets between the template grid and the signal parameters.

This is the basic reason why in this paper we often refer to Monte Carlo studies. In all these studies the choice of signal parameters $\vec\lambda_0$ represents our target source population, which we assume to be uniformly distributed in spin frequency, log-uniformly distributed in spin-down, with orientation $\cos \iota$  uniformly distributed between $-1$ and $1$, polarization angle $\psi$ uniformly distributed in $|\psi| \leq \pi/4$ and source position uniformly distributed on the sky (uniform in $0\leq\alpha < 2\pi$ and in  $-1\leq \sin\delta \leq1$). The log-uniform distribution of spin-down values reflects our ignorance of the actual spin-down distribution of the sources over our large target range.

The Monte Carlo studies make the results robust and simple to interpret: All systematic effects in the analysis, both known and unknown, are automatically incorporated.

We note that since this analysis was carried out, a new metric ansatz was suggested \citep{Allen:2019vcl}, which shows that the metric mismatch generically overestimates the actual mismatch, and shows how to extend the range of validity of the metric approximation. This might mitigate the need for such extensive Monte Carlo studies.

\section{The Einstein@Home search}
\label{sec:stage0}

\subsection{The distribution of the computational load on \EaH}
\label{sub:loadOnEaH}

This search leverages the computing power of the \EaH~project. This is built upon the BOINC (Berkeley Open Infrastructure for Network Computing) architecture~\citep{Boinc1,Boinc2,Boinc3}: a system that uses the idle time on volunteer computers to solve scientific problems that require large amounts of computing power. 

The total number of templates that we searched with \EaH\ is \paramtotaltemplates. The search is split into work-units (WUs) sized to keep the average \EaH\ volunteer computer busy for about {\paramWUcputimeHours} CPU-hours.  A total of {\paramtotalWUsmillions} million WUs are necessary to cover the entire parameter space, representing of order 10 000 CPU-years of computing.

Each WU searches {\paramWUtotaltemplates} templates, and covers $50$ mHz, the entire spindown range and a portion of the sky. Out of the detection statistic values computed for the {\paramWUtotaltemplates} templates, the WU-search returns to the \EaH\ server only the information of the highest {\paramNCand} $\BSNtsc$ results.

This search ran on \EaH\ between April 2018 and July 2019, with an interruption of 8 months at the request of the LIGO/Virgo Collaboration, after the authors left the Collaboration.

\subsection{Post-processing of the \EaH\ search}

\begin{figure*}[h!tbp] 
	\centering
	\includegraphics[width=2 \columnwidth]{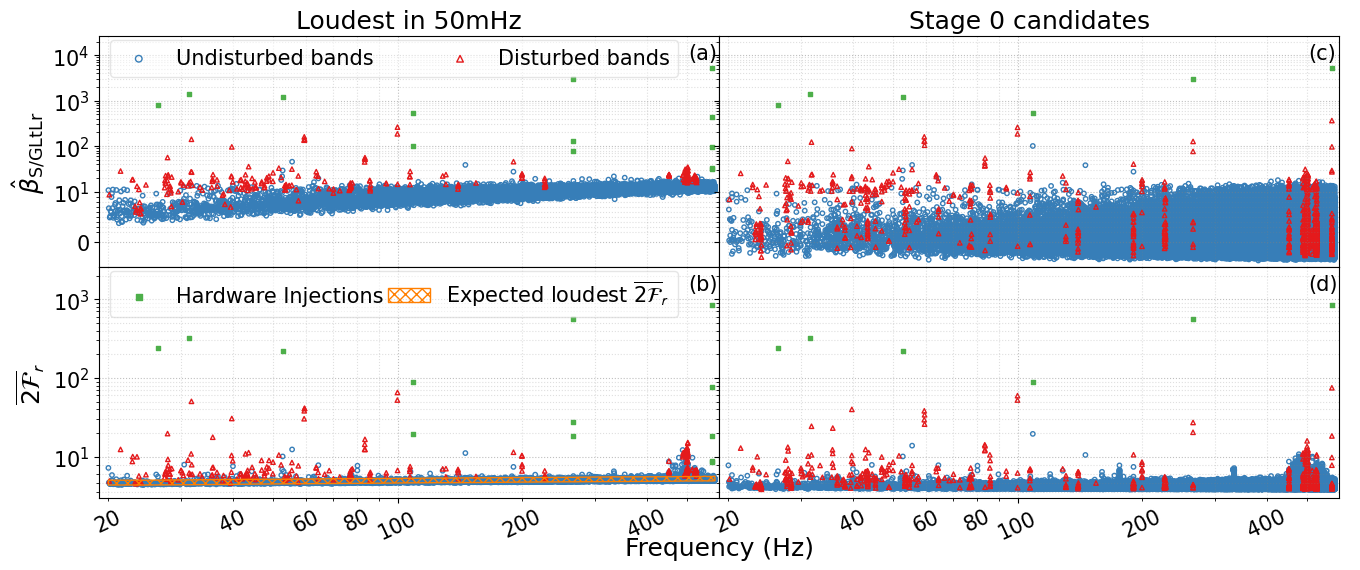}
	\caption{Detection statistics values of candidates as a function of frequency. The candidates coming from undisturbed bands are blue circles, from disturbed bands are red triangles and those from hardware injections are green squares. An unconventional vertical scale is used in all plots, which is linear below  $10$ and log$_{10}$ elsewhere. {\it{Left panels:}} $\BSNtsc{}_r$ and $ 2\avF_r$ value of the loudest candidate (the candidate with the highest $\BSNtsc{}_r$) over 50 mHz, the entire sky and the full spin-down range, out of the \EaH\ search. The increase in detection statistics with frequency is due to the number of searched templates increasing with frequency, as shown in Fig.~\ref{fig:NumberOfTemplatesIn50mHz}. The orange gridded area in the lower left panel indicates the 3$\sigma$ expected range  in Gaussian noise. {\it{Right panels:}} Detection statistics values of the \NCandFUZero\ candidates that are followed-up.  By comparing the right and left panels one can see how we ``dig'' below the level of the loudest 50-mHz candidate with our follow-up stages.}
	\label{fig:loudestHalfHzAndFU0CandidatesVersusFreq}
\end{figure*}

We refer to a waveform template and the associated search results as a ``candidate".
All in all the \EaH\ search returns \paramNCandTotal\ candidates: the top \paramNCand\ candidates per WU  $\times ~{\paramtotalWUsmillions}$ million WUs.  This is where the post-processing begins. 

The post-processing consists of three steps:
\begin{itemize}
\item {\bf{Banding:}} as described in the previous section, each volunteer computer searches for signals with frequency within a given 50-mHz band, with spin-down between $\paramfdotmin$ and $\paramfdotmax$ Hz/s and a portion of the sky. The first step of the post-processing is to gather together all results that pertain to the same 50-mHz band. We compute some basic statistics from these results and produce a series of diagnostic plots, that we can conveniently access through a GUI (graphical user interface) tool that we have developed for this purpose. This provides an overview of the result-set in any 50-mHz band. 

\item {\bf{ Identification of disturbed bands:}} as done in previous \EaH\ searches \citep{Papa:2020vfz,Ming:2019xse,Abbott:2017pqa,Zhu:2016ghk,Singh:2016phs} we identify bands that present very significant deviations of the detection statistics from what we expect from a reasonably clean noise background. Such deviations can arise due to spectral disturbances or to extremely loud signals. We do not exclude these bands from further inspection, but we do flag them as this information is necessary when we set upper limits. We mark \NDisturbedBands\ 50-mHz bands as disturbed.

\item {\bf{ Clustering:}} in this step we identify clusters of candidates that are close enough in parameter space that they are likely due to the same root cause. We associate with each cluster the template values of the candidate with the highest $\BSNtsc{}_r$, which we also refer to as cluster seed. 
We use a new clustering method \citep{densityclustering} that identifies regions in frequency-spin-down-sky-position that harbour an over-density of candidates -- a typical signal signature. This method achieves a lower false dismissal of signals at fixed false alarm rate, with respect to the previous clustering \citep{Singh:2017kss} by tracing the SNR reduction function with no assumption on its profile in parameter space. An occupancy veto is also applied, requiring at least 4 candidates to be associated with a cluster. Most candidates have {\it fewer} than three nearby partners, so this clustering procedure greatly reduces the number of candidates, namely from  $\paramNCandTotal\ $ to \NCandFUZero.

\item {\bf{Follow-up searches:}} After the clustering we have \NCandFUZero\ candidates, shown in Figure \ref{fig:loudestHalfHzAndFU0CandidatesVersusFreq}. Of these, \NCandDisturbedFUZero\ come from bands that have been marked as disturbed. We follow all of them up as detailed in the next Section. The list of the disturbed 50-mHz bands is provided in \citet[and Suppl. Mat.]{O2AS-AEI}.

\end{itemize}

In order to give a sense of the overall set of \EaH\ results, in the left panels of Figure \ref{fig:loudestHalfHzAndFU0CandidatesVersusFreq} we show the detection statistic value of the most significant result from every 50-mHz band. 
The large majority of the results falls within the expected range for noise-only. Most of the highest detection statistic values stem from hardware injections or from disturbed bands and are due to spectral contamination, i.e. signals (as opposed to noise fluctuations) of non-astrophysical origin.

\begin{deluxetable*}{lcccccccccccc}
\tablecaption{Overview of the searches. We show the values of the following parameters: the number of segments $\Nseg$ and the coherent time baseline of each segment $T_\mathrm{coh}$; the grid spacings $\delta f,\delta\fdot$ and $m_{\text{sky}}$; the average mismatch $<\mu >$; the parameter space volume searched around each candidate, $\pm\Delta f, \pm\Delta\fdot$ and ${\textrm{r}_\textrm{sky}}$ expressed in units of the side of the hexagon sky-grid tile of the Stage 0 search (Eq.~\ref{eq:skyGridSpacing}); the threshold value $R^{a}$ used to veto candidates of Stage $a$ (Eq.~\ref{eq:R}); the number of templates searched (N$_{\textrm{in}}$ ) and how many of those survive and make it to the next stage (N$_{\textrm{out}}$). The first search, Stage 0, is the \EaH\ search, hence the searched volume is the entire parameter space. The other searches are the follow-up stages. \label{tab:FUtable}
}
\tablehead{
\colhead{Search} & \colhead{$T_\mathrm{coh}$} & \colhead{$\Nseg$} & \colhead{$\delta f$} & \colhead{$\delta {\dot{f}}$} &  \colhead{$m_{\text{sky}}$} & \colhead{$< \mu >$ } & \colhead{$\Delta f$} & \colhead{$\Delta\fdot$} & \colhead{$ {\textrm{r}_\textrm{sky}\over {d(\mskyFUZero )}}$} & \colhead{$R^{a}$} & \colhead{ N$_{\textrm{in}}$ }& \colhead{N$_{\textrm{out}}$} \\
& hr &  & $\mu{\textrm{Hz}}$ &  {{ $10^{\scriptstyle{-14}}$} Hz/s}  &  & & $\mu{\textrm{Hz}}$ & {{ $10^{\scriptstyle{-14}}$} Hz/s} &  &   &  &
%
}
\startdata
%
\TBstrut Stage 0 & $\TcohFUZero$ 	& $\NSegFUZero$ 	& $\dfreqmuHzFUZero$ 	& $\dfdotfgTenfHzFUZero$ 	&   $\mskyFUZero$ & \meanMismatchFUZero & \tiny{full range} 	& \tiny{full range}	& \tiny{all-sky} 	&  $\rVetoFUZero$	& $\paramtotaltemplates$  & $\NCandFUZero$ \\
\TBstrut Stage 1 & $\TcohFUZeroR$ 	& $\NSegFUZeroR$ 	&  $\dfreqmuHzFUZeroR$ 	& $\dfdotfgTenfHzFUZeroR$	&  $\mskyFUZeroR$  & \meanMismatchFUZeroR & $\CRfreqmuHzFUZero$ 	& $\CRfdotFUZero$ 	& $\CRskyFUZero$ 	& $\rVetoFUZeroR$ 	&   $\NCandFUZero$ & \NCandFUZeroR  \\
\TBstrut Stage 2 & $\TcohFUOne$ 		& $\NSegFUOne$ 	& $\dfreqmuHzFUOne$ 	& $\dfdotfgTenfHzFUOne$ 	& $\mskyFUOne$  & \meanMismatchFUOne & $\CRfreqmuHzFUZeroR$ 	& $\CRfdotFUZeroR$ 	&  $\CRskyFUZeroR$ 	& $\rVetoFUOne$ 	&  $\NCandFUZeroR$  & \NCandFUOne\\
\TBstrut Stage 3 & $ \TcohFUOneR$ 	& $\NSegFUOneR$ 	& $\dfreqmuHzFUOneR$ 	& $\dfdotfgTenfHzFUOneR$  	& $\mskyFUOneR$  & \meanMismatchFUOneR  & $\CRfreqmuHzFUOne$ 		& $\CRfdotFUOne$ 	& $\CRskyFUOne$ 	& $\rVetoFUOneR$ 	& $\NCandFUOne$  & \NCandFUOneR \\
\TBstrut Stage 4 & $\TcohFUTwo$ 		& $\NSegFUTwo$	& $\dfreqmuHzFUTwo$ 	& $\dfdotfgTenfHzFUTwo$  	& $\mskyFUTwo$  & \meanMismatchFUTwo  & $ \CRfreqmuHzFUOneR$ 	& $\CRfdotFUOneR$ 	& $\CRskyFUOneR$ 	& $\rVetoFUTwo$  	& $\NCandFUOneR$  & \NCandFUTwo\\
\TBstrut Stage 5 & $\TcohFUThree$	& $\NSegFUThree$ 	& $\dfreqmuHzFUThree$ 	&  $\dfdotfgTenfHzFUThree$	& $\mskyFUThree$  & \meanMismatchFUThree & $\CRfreqmuHzFUTwo$ 		& $\CRfdotFUTwo$	& $\CRskyFUTwo$ 	& $\rVetoFUThree$  	& $\NCandFUTwo$  & \NCandFUThree \\
\TBstrut Stage 6 & $\TcohFUFour$ 	& $\NSegFUFour$ 	& $\dfreqmuHzFUFour$ 	&$\dfdotfgTenfHzFUFour$ 	& $\mskyFUFour$  & \meanMismatchFUFour & $\CRfreqmuHzFUThree$	& $\CRfdotFUThree$ 	& $\CRskyFUThree$ 	&  $\rVetoFUFour$	& $\NCandFUThree$  & \NCandFUFour \\
\TBstrut Stage 7 & $\TcohFUFive$ 		& $\NSegFUFive$ 	& $\dfreqmuHzFUFive$  	& $\dfdotfgTenfHzFUFive$  	&  $\mskyFUFive$  & \meanMismatchFUFive & $\CRfreqmuHzFUFour$ 	& $\CRfdotFUFour$ 	& $\CRskyFUFour$ 	&$\rVetoFUFive$ 	& $\NCandFUFour$  & \NCandFUFive \\
\TBstrut Stage 8 & $\TcohFUSix$			& $\NSegFUSix$ 	& \dfreqmuHzFUSix 	& \dfdotfgTenfHzFUSix	& \mskyFUSix & \meanMismatchFUSix & $\CRfreqmuHzFUFive$ 		& $\CRfdotFUFive$ 	& $\CRskyFUFive$  	& $\rVetoFUSix$  	&  $\NCandFUFive$  & 6 \\
\enddata
\end{deluxetable*}

\section{The follow-up searches}
\label{sec:followups}

Each stage takes as input the candidates that have survived the previous stage. Waveforms around the nominal candidate parameters are searched, so that if the candidate were due to a signal it would not be missed in the follow-up. The extent of the volume to search is based on the results of injection-and-recovery Monte Carlo studies and is broad enough to contain the true signal parameters for $\gtrsim 99.8\%$ of the signal population. For this reason we also refer to this volume as the ``signal-containment region"\footnote{The Monte Carlos were performed with \num{1839} signals, of which in Stage 1 the chosen containment region contained \num{1836}. For the other stages all the signals were recovered within the chosen containment regions.}. The containment region in the sky is a circle in the orthogonally projected ecliptic plane with radius $\textrm{r}_\textrm{sky}$.

\begin{figure}[h!tbp] 
\includegraphics[width=\columnwidth]{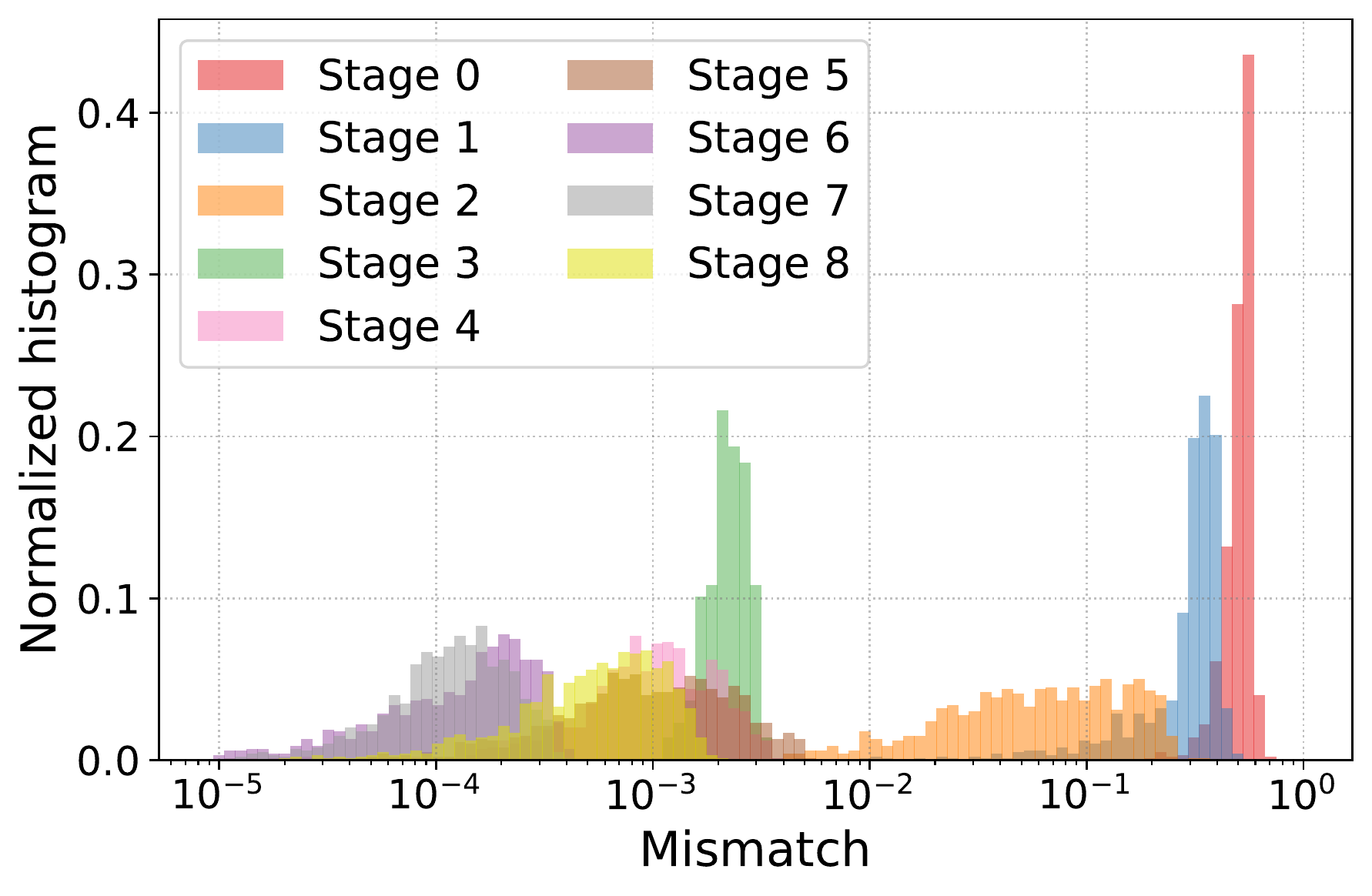}
        \caption{Mismatch distributions for the various follow-up searches based on \num{1000} injection-and-recovery Monte Carlos. The search set-ups are chosen so that the SNR of a signal increases from one stage to the next. This is achieved either by increasing the $\Tcoh$ of the search and/or by decreasing the mismatch. We note that even though the average mismatch of Stage 8 is larger than that of the previous two stages, this does not imply that the expected SNR for a signal out of Stage 8 is smaller.}
\label{fig:mismatches}
\end{figure}
The search set-ups for Stages 1-8 are chosen so that the SNR of a signal would increase from one stage to the next. This is achieved in two ways: by increasing the $\Tcoh$ of the search and/or by using a finer grid and hence by decreasing the average mismatch. The mismatch distributions of the various searches are shown in Figure \ref{fig:mismatches}. We note that even though average mismatch of Stage 8 is larger than that of the previous two stages, this does not imply that the expected SNR for a signal out of Stage 8 is smaller. In fact, because of the larger $\Tcoh$ used in Stage 8, the expected SNR for a signal out of Stage 8 is larger than that of the same signal out of Stage 7 or 6. This can be seen by comparing the values of $R^8$,  $R^7$ and $R^6$, in Table \ref{tab:FUtable} (the quantity $R^a$ is defined below in Eq.~\ref{eq:R} and is related to the expected SNR increase at Stage $a$ with respect to Stage $0$).

We cluster the results of each search and consider the most significant cluster. 
We associate to the cluster the parameters of the member with the highest detection statistic value, and refer to this as the candidate from that follow-up stage.

\begin{figure}
	\includegraphics[width=\columnwidth]{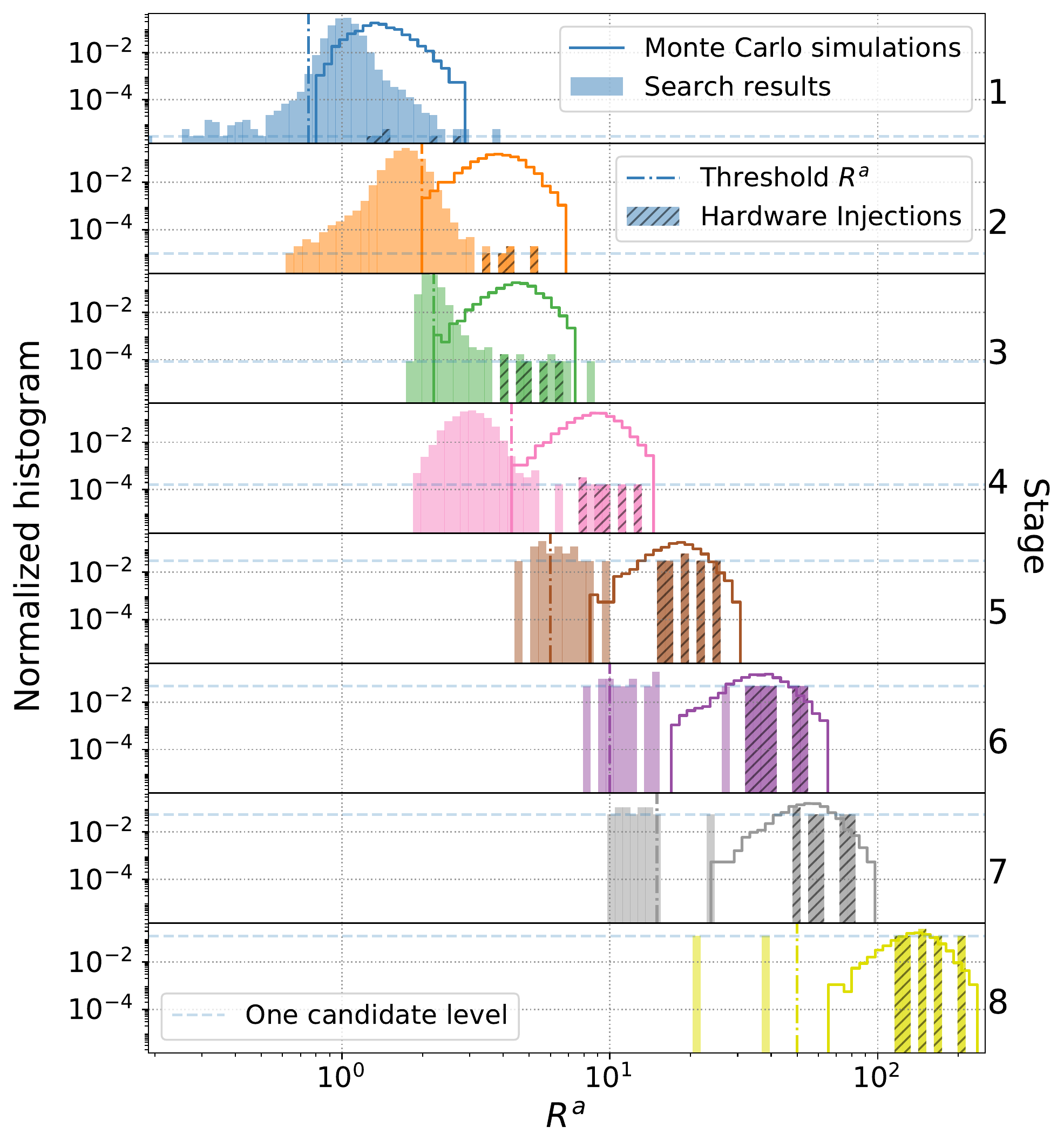}
	\caption{Distributions of $R^a$ of candidates from signal injection-and-recovery Monte Carlos (solid lines) and from the actual search (shaded areas). The dashed-shaded areas show the $R^a$ bins associated with the hardware injections. The dashed vertical lines mark the $R^{a}$ threshold values. 
	The dashed horizontal lines mark the 1-candidate level in the search results.}
	\label{fig:RVeto}
\end{figure}

We veto candidates at stage $a$ whose SNR does not increase as expected for signals, with respect to Stage 0. We do this by setting a threshold on the quantity 
\begin{equation}
R^{a}={2\avF_r^{\textrm{~Stage a}} - 4 \over {2\avF_r^{\textrm{~Stage 0}} -4}}.
\label{eq:R}
\end{equation}
The threshold is set based on signal injection-and-recovery Monte Carlos, as shown in Figure \ref{fig:RVeto}. The values are given in Table \ref{tab:FUtable}. Because of the large number of candidates in the first four follow-up stages, the $R^a$ thresholds for $a=1\cdots 4$ are stricter than those used for the last four stages. 

All the parameters relative to the searches, as well as the number of candidates surviving each stage, are shown in Table \ref{tab:FUtable}. 

Only six candidates are left at the output of Stage 8. They are due to fake signals present in the data stream for validation purposes, the so-called hardware-injections. In fact there are six hardware injections with parameters that fall in our search volume, those with ID 0, 2, 3, 5, 10, 11 \citep{O2_injection_params}. We recover them all with consistent parameters.

\section{Upper limits}
\label{sec:ULs}

\begin{figure}
\includegraphics[width=\columnwidth]{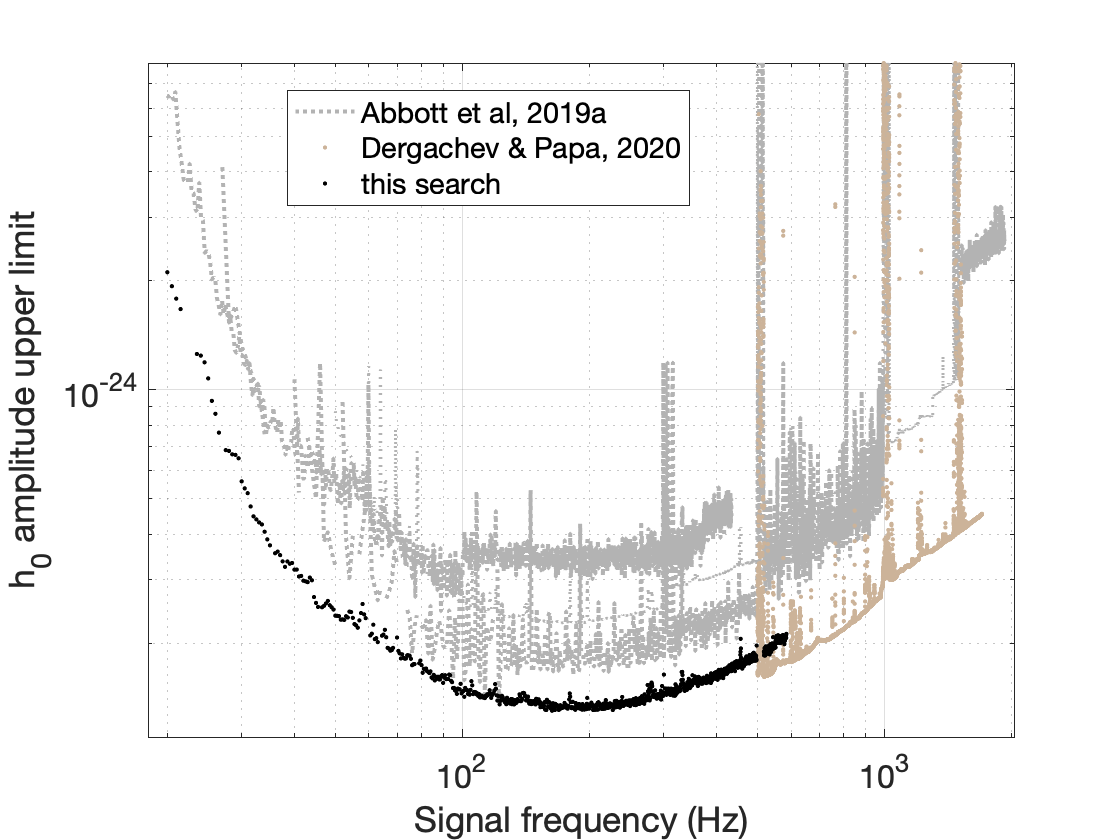}
\caption{Smallest gravitational wave amplitude $h_0$ that we can exclude from the assumed population of signals (see Section \ref{sub:MCs}). 
We compare our results with the latest literature: the Falcon search \citep{Dergachev:2020fli} and the LIGO results \citep{Pisarski:2019vxw} on the same data. There are multiple curves associated with the LIGO results because they used different analysis pipelines.}
\label{fig:h0ULs}
\end{figure}

Based on our null result we set 
$90\%$ confidence 
frequentist upper limits on the gravitational wave amplitude $h_0$ in half-Hz bands. The upper limit value is the smallest signal amplitude that would have produced a signal above the sensitivity level of our search for 90\% of the signals of our target population (see Section \ref{sub:MCs}). We establish the detectability of signals based on injection-and-recovery Monte Carlos. The upper limits are shown in Figure \ref{fig:h0ULs} and provided in machine-readable format in \citet[and Suppl. Mat.]{O2AS-AEI}.

Our upper limits do not hold in some 50-mHz bands, namely those marked as disturbed and those associated with the hardware injections. Even though we have followed-up candidates from these bands, we cannot exclude that a signal with strength below the disturbance but above the detection threshold -- and hence above the upper limit -- could be hidden by the loud disturbance, for example by being associated with its large noise-cluster. Another reason why we cannot guarantee that our upper limit holds in the presence of a disturbance is the saturation in the \EaH\ top-list that a loud disturbance produces. This prevents candidates from quieter parameter space regions in that band from being recorded. Given how loud the hardware injections are, for similar reasons, we also exclude the 50-mHz bands associated with these. The 50-mHz bands where the upper limits do not hold are provided in \citet[and Suppl. Mat.]{O2AS-AEI}. 

Upper limits are also not given in some half-Hz bands. This happens for two reasons: 1) If all 50-mHz bands in a half-Hz band are disturbed 2) due to the bin-cleaning procedure: In Section \ref{subsec:data} we explained that we remove contaminated frequency bins and substitute them with Gaussian noise. If a signal were present in the cleaned-out bins, it too, would be removed. So in the half-Hz bands affected by cleaning, the upper limit Monte Carlos include the cleaning step {\it{after}} the signal has been added to the data. In this way the loss in detection efficiency due to the cleaning procedure is naturally folded into the upper limit. When a large fraction of the half-Hz bins is cleaned out, however, the detection efficiency may not reach the target 90\% level. In this case we do not give an upper limit in the affected band. 
The list of half-Hz bands for which we do not give upper limits is given in \citet[and in the Suppl. Mat.]{O2AS-AEI}. 

Based on the upper limits, we compute the sensitivity depth $\D$ of the search \citet{Behnke:2014tma} and find values between (\minSensitivityDepth\ - \maxSensitivityDepth) 
$1/\sqrt{\textrm{Hz}}$. This is consistent with, and slightly better than, previous performance of \EaH\ searches \citep{Dreissigacker:2018afk}. We provide the power spectral density estimate used to derive the sensitivity depth in \citet[and in the Suppl. Mat.]{O2AS-AEI}.

\begin{figure}
\includegraphics[width=\columnwidth]{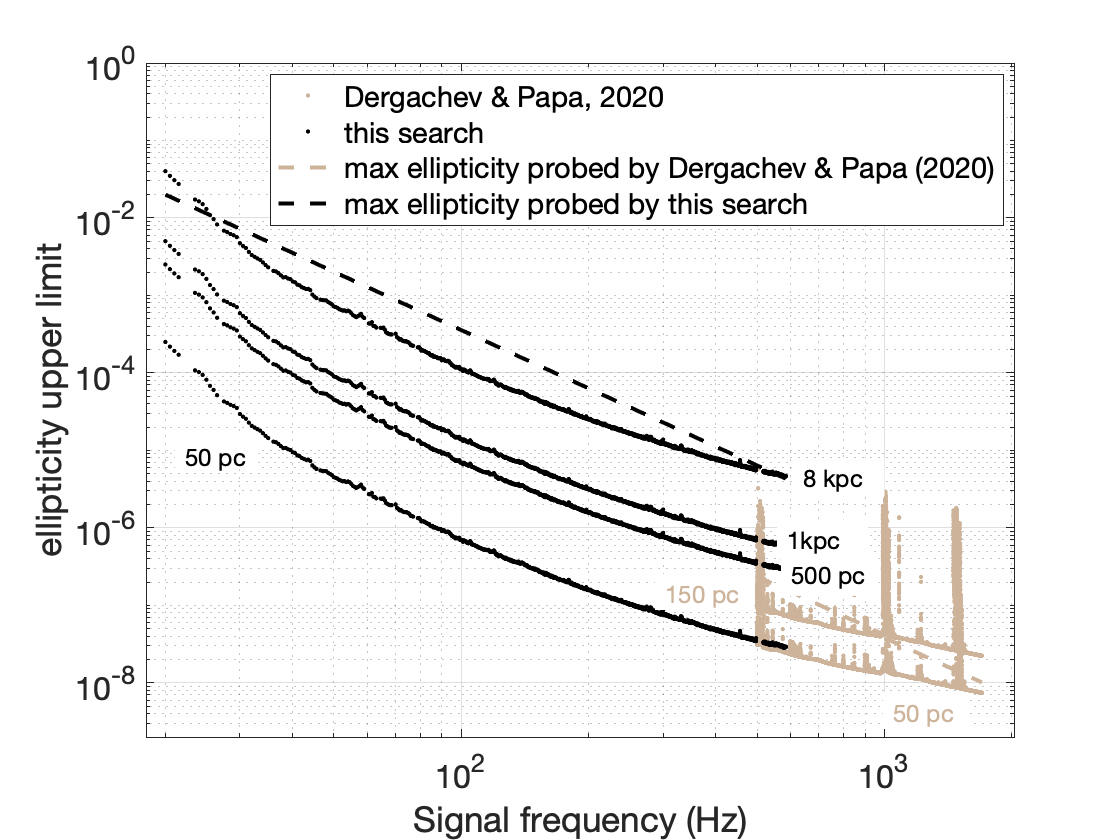}
\caption{Upper limits on the ellipticity of a source at a certain distance (black). We also show the recent upper limits from the low ellipticity all-sky search of \cite{Dergachev:2020fli}. The dashed line is the spin-down ellipticity for the highest spin-down rate probed by each search. }
\label{fig:epsilonULs}
\end{figure}

We can express the $h_0$ upper limits as upper limits on the ellipticity $\varepsilon$ of a source modelled as a triaxial ellipsoid spinning around a principal moment of inertia axis $\hat{I}$ 
at a distance $D$ \citep{Jaranowski:1998qm, Gao:2020zcd}:
\begin{equation}
\begin{split}
\varepsilon &=1.4 \times 10^{-6} ~\left( {h_0\over{1.4\times10^{-25}}}\right ) \times \\
&\left ( {D\over{1~\textrm{kpc}}}\right ) \left ({{\textrm{170~Hz}}\over f} \right )^2 \left ({10^{38}~{\textrm{kg m}}^2\over I} \right ).\\
\end{split}
\label{eq:epsilon}
\end{equation}
The ellipticity $\varepsilon$ upper limits are plotted in Figure \ref{fig:epsilonULs}. If the spin-down of the signal were just due to the decreasing spin rate of the neutron star, then our search could not probe ellipticities higher than the spin-down limit ellipticity corresponding to the highest spin-down rate considered in the search, $\paramfdotmin$ Hz/s.  This is indicated in Figure \ref{fig:epsilonULs} by a dashed line. 

Proper motion can reduce the  apparent spin-down \citep{Shklovskii1970}, so in principle we could detect a signal from a source with ellipticity above the dashed line. However, even in extreme cases (source distance 8 kpc, spin period 1 ms, large proper motion 100 mas/yr \citep{Hobbs:2005yx} or source distance 10 pc, spin period 1 ms and tangential velocity of 1000 km/s ) the change in maximum detectable ellipticity is negligible.

\section{Conclusions}
\label{sec:conclusions}

We present the results from an \EaH\ search for continuous, nearly monochromatic, gravitational waves with frequency between $\paramfmin$ and $\paramfmax$ Hz, and spin-down between $\paramfdotmin$ and $\paramfdotmax$ Hz/s. We use LIGO O2 public data and compare it against $\paramtotaltemplates\ $ waveforms. We follow-up the most likely \NCandFUZero\ candidates through a hierarchy of eight searches, each being more sensitive but requiring more per-template computing power than the previous one. No candidate survives all the stages. 

This is the most sensitive search performed on this parameter space on O2 data, and sets the most stringent upper limits on the intrinsic gravitational wave amplitude $h_0$. The most constraining $h_0$ upper limit is $\mostStringentUL\ $ at $\mostStringentULFreq\ $ Hz, corresponding to a neutron star at, say, 100 pc, having an ellipticity of $\lesssim 5\times 10^{-7}$ and rotating with a spin period of $\approx$ 12 ms. Our results thus exclude neutron stars rotating faster than 12 ms, within 100 pc of Earth, with ellipticities in the few $\times 10^{-7}$ range and reach the $1\times 10^{-7}$ mark for spins of 5 ms. 

These results probe a plausible range of pulsar ellipticity values, well within the boundaries of what the crust of a standard neutron star could support , around $10^{-5}$, according to some models \citep{JohnsonMcDaniel:2012wg}. It is hard to produce a definitive estimate of such quantity and it may be that this maximum value is significantly lower \citep{Gittins:2020cvx}. Since the closest neutron star is expected to be at about a distance of 10 pc \citep{Dergachev:2020fli}, it is likely that there are several hundreds within 100 pc. On the other hand, recent analyses of the population of known pulsars suggest that their ellipticity should lie in the $10^{-9}$ decade \citep{Woan:2018tey,Bhattacharyya:2020paf}, which we reach only for sources rotating faster than 5 ms and within 10 pc. When the O3 LIGO data is released, its sensitivity improvement with respect to the O2 data used here \citep{Buikema:2020dlj} will allow us to extend the reach of our search and probe ellipticities in the $10^{-9}$ decade, at these higher frequencies.
 

\acknowledgments
We thank the \EaH\ volunteers, without whose support this search could not have happened. \\
We acknowledge the NSF grant Nr.1816904. \\
The follow-up searches were all performed on the ATLAS cluster at AEI Hannover. We thank Carsten Aulbert and Henning Fehrmann for their support. \\
This research has made use of data, software and/or web tools obtained from the LIGO Open Science Center (\url{https://losc.ligo.org}), a service of LIGO Laboratory, the LIGO Scientific Collaboration and the Virgo Collaboration.  LIGO is funded by the U.S. National Science Foundation. Virgo is funded by the French Centre National de Recherche Scientifique (CNRS), the Italian Istituto Nazionale della Fisica Nucleare (INFN) and the Dutch Nikhef, with contributions by Polish and Hungarian institutes.

\newpage

\bibliography{paperBibApJ}{}
\bibliographystyle{aasjournal}
\bibstyle{aasjournal}

\end{document}